\documentstyle[12pt,epsf]{article}

\newcommand{\be}{\begin{equation}}
\newcommand{\ee}{\end{equation}}
\newcommand{\ba}{\begin{eqnarray}}
\newcommand{\ea}{\end{eqnarray}}
\newcommand{\simm}[1]{\stackrel{#1}{\scriptstyle\sim}}

\def\betap{\beta_{+}}
\def\betam{\beta_{-}}
\def\NtFourLattice{$18^2 \times 48 \times 4$}
\def\omrot{\Omega_{\rm rot}}

\begin{document}

\begin{titlepage}
\hfill UTHEP-257 \\

\begin{centering}
\vfill
{\bf Interface tension in SU(3) lattice gauge theory at finite temperatures on
an $N_t=2$ lattice}

\vspace{1cm}
Yasumichi Aoki and Kazuyuki Kanaya

\vspace{1cm}
{\em
Institute of Physics, University of Tsukuba, 

Tsukuba, Ibaraki 305, Japan
}

\end{centering}

\vspace{1.5cm}\noindent
The surface tension $\sigma$ of the confined-deconfined interface is
calculated in pure $SU(3)$ lattice gauge theory at finite temperatures
employing  the operator and integral methods on  a lattice of a size $8^2\times
N_z\times 2$ with $N_z=16$ and 40.  Analyses of non-perturbative corrections
in asymmetry response functions strongly indicate that the use of
one-loop values for the response functions lead to an overestimate of
$\sigma$ in the operator method.  The operator method also suffers more
from finite-size effects due to a finite  thickness of the interface, leading
us to conclude that the integral method yields  more reliable values for
$\sigma$.  Our result with the integral method $\sigma/T_c^3=0.134(16)$ 
is consistent with earlier results and also with that obtained with a 
transfer matrix method. 
Result is also reported on $\sigma$ obtained on a lattice
$18^2\times 48\times 4$ with the integral method. \vfill \noindent

\vfill \noindent

\end{titlepage}


%%%%%%%%%%%%%%%%%%%%%%%%%%%%%%%%%%%%%%%%%%%%%%%%%%%%%%%%%%%%%%%%
\section{Introduction}

Numerical simulation of pure $SU(3)$ gauge system on a lattice has shown
that the system undergoes a first order phase transition from a
confined phase at low temperatures to a deconfined phase at high temperatures
\cite{review}. The two phases can therefore coexist  at the
transition temperature $T_c$, separated by an interface.  

A basic parameter characterizing
the interface is the interface tension $\sigma$. A number of numerical work
has recently been carried out to determine its value, mainly for a system
with the temporal lattice size $N_t=2$.     Kajantie,
K\"arkk\"ainen and Rummukainen\cite{Helsinki88,Helsinki90}  developed an
operator method for measuring the tension and reported the value 
$\sigma/T_c^3 = 0.24(6)$ for  $N_t=2$\cite{Helsinki90}.  
Independently, Potvin and Rebbi\cite{Boston89} proposed to employ an
integral of the derivative of the free energy in the parameter space of the
coupling constant to measure the tension, and found
$\sigma/T_c^3 = 0.115(13)$ for the same temporal lattice size\cite{Boston90}. 
A factor two discrepancy between these results has motivated
further studies of the
interface\cite{Grossmann92b,Berg92,Grossmann92,Grossmann93}, which yielded 
values of $\sigma/T_c^3$ similar to that of the integral method.  
The method of histograms\cite{binder}
%\cite{binder,Wiese92,Bunk92,Caselle} 
and the technique of transfer
matrix\cite{Grossmann92b} used in these studies are quite different, however,
from the operator and the integral method.  Hence the discrepancy of the
original results obtained with the two methods has not been resolved.

In this article we report on our attempt toward an understanding of the origin
of the discrepancy through a comparative study of the two methods.  
Since a
possible cause of the discrepancy is the use of a different spatial lattice
size ($8^2\times 40\times 2$ for the operator method\cite{Helsinki90} and  $8^2\times 16\times
2$ for the integral method\cite{Boston90} ),  we carry out
simulations for both sizes of lattice with both methods.  
We also increase statistics by a factor of two to four over those of
ref.~\cite{Helsinki90,Boston90}.
We find that a part of the discrepancy can be understood by the effect 
of the finite thickness of the interface. 
Comparing the results on a large lattice with $N_z=40$, where this 
problem is absent, we find, however 
that the discrepancy is not completely removed for $N_t=2$. 

A nice feature of the integral method is that it directly relates plaquette
expectation values to the interface tension and  no other input is needed.   
In contrast the operator method requires the values
of response functions of the coupling constant under an anisotropic deformation
of lattice.  Up to now they have been estimated only in one-loop 
perturbation theory\cite{karsch82}, whose applicability  may be questioned for
small values of the inverse gauge coupling constant $\beta=6/g^2$, such as 
the critical value $\beta_c$ for $N_t=2$.  
We discuss the effect of nonperturbative corrections to these coefficients. 

Another problem which became apparent in recent work on the interface 
tension is the difficulty of measuring its value for lattices with
the temporal size $N_t\geq 4$.  For the case of $N_t=4$ an initial attempt with
the integral method failed to obtain a non-vanishing value\cite{Boston90}.   An
alternative integral method employing an external field coupled to the Polyakov
loop was then devised, and  a value $\sigma/T_c=0.027(4)$ has been
obtained\cite{Brower92}.   We attempt a high statistics measurement of
the interface tension with the original integral method on a $18^2\times
48\times 4$ lattice. The results will also be reported in this article.

This paper is organized as follows.  
In Sec.~2 we briefly recall the operator and the integral 
method and describe our simulations.  
Our results for the interface tension are presented in 
Sects.~\ref{Results2} and \ref{Results4}  for $N_t=2$ and 4, respectively.
Sec.~\ref{Conclusion} is devoted to conclusions.

%%%%%%%%%%%%%%%%%%%%%%%%%%%%%%%%%%%%%%%%%%%%%%%%%%%%%%%%%%%%%%%%
\section{Method and simulation}		\label{method}
We consider the pure $SU(3)$ gauge system with the standard plaquette action on a
lattice of a size $N_x\times N_y\times N_z\times N_t$ with the size $N_z$ in
the $z$-direction at least twice as large as those in the $x$- and
$y$-directions.  If one splits the lattice into two halves in the
$z$-direction and chooses $\beta$ in one half of the
lattice to be slightly below the critical value
$\beta_1=\beta_- \equiv \beta_c-\delta\beta$ and that of the other 
half slightly above
$\beta_2=\beta_+ \equiv \beta_c+\delta\beta$, 
the system in the first half will be in
the confined phase and the other half in the deconfined phase.  With the
periodic boundary condition employed for the gauge field in all four 
directions,
there will be two interfaces in the $x-y$ plane, each with a transverse area 
$A=N_xN_ya^2$ with $a$ the lattice spacing.

The operator method\cite{Helsinki88,Helsinki90} starts from the
following thermodynamic relation  defining the interface tension $\sigma$ in
terms of the free energy $F$,  
\be
 \sigma 
 = \left. \frac{1}{2}\frac{\partial F}{\partial A} \right|_{T,V}\ 
\label{eq:dFdAop}
\ee
where the factor $1/2$ is to take into account the presence of two interfaces.  
For the standard plaquette action, working out
the derivative yields
\cite{Helsinki90},
\be
 \frac{\sigma A}{T} = 
 \frac{1}{2}\sum_n [\beta_n + 3(c_t(\beta_n) - c_s(\beta_n)]
 (2P_{n;12} - P_{n;13} - P_{n;23} 
   - 2P_{n;03} + P_{n;01} + P_{n;02})
\label{eq:operator}
\ee
where $T=1/(N_ta)$ is the temperature, $\beta_n$ is the local value of $\beta$ at
site $n$ taking either of the values $\beta_1=\beta_-$ or $\beta_2=\beta_+$,
and  $
  P_{n;\mu\nu} 
  = \langle \frac{1}{3} {\rm Re Tr} U_{n;\mu\nu}\rangle
$
is the plaquette expectation value in the $(\mu,\nu)$ plane at site $n$.  The
functions $c_s(\beta)$ and $c_t(\beta)$ are the response functions of  the
gauge coupling constant  with respect to an anisotropic deformation of the
lattice  in the spatial and temporal directions: 
$c_i = (\partial g^{-2}_i / \partial \xi)_{\xi=1}$, $i=s$, $t$,
and $\xi = a_s/ a_t$. 
In perturbation theory these functions are expanded as\cite{karsch82} 
\ba
c_s(\beta) &=& 0.20161+O(\beta^{-1})\ ,\cr 
c_t(\beta) &=& -0.13195+O(\beta^{-1})\ .
\label{eq:asymmetry}
\ea 

The integral method \cite{Boston89,Boston90} is based on the relation 
\ba
 \sigma A &  =  &\frac{1}{2}\big[ F(\betam,\betap) 
   - \frac{1}{2} \{F(\betap,\betap)) + F(\betam,\betam)\}\big] \cr
 &  = & \frac{1}{4} [ \{F(\betam,\betap) - F(\betam,\betam)\}
              - \{F(\betap,\betap) - F(\betam,\betap)\} ]
\label{eq:integralmethod}
\ea
where $F(\beta_1,\beta_2)$ is the free energy of the system with the two halves
having a coupling  $\beta_1$ and $\beta_2$.  Denoting expectation values for
such a system as $\langle \cdot \rangle_{\beta_1 \beta_2}$ and decomposing the
total action into the contributions of the two halves  $S = \beta_1 S_1 +
\beta_2 S_2$, we have  
\be
  \frac{1}{T} \frac{\partial F(\beta_1,\beta_2)}{\partial \beta_i} 
  = - \langle S_i \rangle_{\beta_1\beta_2}\ ,\qquad i=1,2
\label{eq:dFdB}
\ee 
Expressing the free energy differences in (\ref{eq:integralmethod})
through integrals of (\ref{eq:dFdB}) we obtain
\be
 \frac{\sigma A}{T}  =  \frac{1}{4} [ 
  - \int_{\betam}^{\betap} d\beta_2 \langle S_2 \rangle_{\betam \beta_2}
  + \int_{\betam}^{\betap} d\beta_1 \langle S_1 \rangle_{\beta_1 \betap}
  ]. 
\label{eq:int-path}
\ee

For both the integral and operator methods numerical simulations yield an
estimate of $\sigma$ for a finite difference $\delta\beta$ between the
couplings $\beta_-=\beta_c-\delta\beta$ and $\beta_+=\beta_c+\delta\beta$
chosen for the two halves of the lattice.  The physical value of the interface
tension is obtained by an extrapolation to the limit 
$\delta\beta\rightarrow 0$.

Our simulations are carried out on a lattice of a size $8^2\times N_z\times 2$
with $N_z=16$ and 40, and also on a  $18^2\times 48\times 4$
lattice.  
Gauge configurations are generated through the standard pseudo-heat bath
algorithm with eight hits per link.  

In order to create an interface we take the
gauge coupling constant to be $\beta_1$ for all the
plaquettes starting at the site $(x,y,z,t)$ with $0 \le z < N_z/2$  
and pointing in the positive $\mu\nu$ direction.  
For the rest of plaquettes  at $N_z/2 \le z< N_z$, we assign $\beta_2$.  
Thus the coupling constant jumps 
sharply at $z=0$ and $N_z/2$.    This assignment 
is slightly different from that adopted in ref.~\cite{Helsinki90,Boston90} 
where the lattice is split into two at $z=1/2$ and $N_z/2+1/2$ and the 
coupling $\beta_1$
or $\beta_2$ is used for updating links in the left or right half of the
lattice, while the mean value $(\beta_1+\beta_2)/2$ is chosen for links
connecting $z=0$ to 1 and $N_z/2$ to $N_z/2+1$.   
We prefer our assignment since (i)
thermodynamic  interpretation is clearer with $\beta$ associated to plaquettes 
so that the convergence to thermal equilibrium is guaranteed in numerical
simulations,  and (ii) there is no ambiguity in expressing  $\partial F /
\partial \beta_{i}$  $(i=1,2)$, needed in the integral method,  in terms of
plaquette  expectation values. We note that the physical value of the interface
tension obtained in the limit $\beta_1-\beta_2\rightarrow 0$ should not depend
on the details of assignment.   This is found to be the case as will be 
shown below.

Another choice one has to make in applying the two methods is the value of the
critical coupling $\beta_c$, which involves ambiguities on a finite lattice. 
A possible definition of
$\beta_c$ on a finite lattice is  the peak position of the susceptibility of 
the $Z(3)$-rotated Polyakov loop.  Carrying out runs on a  $8^2\times N_z\times
2$ lattice with $\beta_1=\beta_2$ and  applying the spectral density
method\cite{swend} to locate the peak position,  we find that $\beta_c$=
5.09146(63) and 5.09409(38) for $N_z=16$ and 40.  These values are
slightly higher than the choice $\beta_c=5.09$ used in
refs.~\cite{Helsinki90,Boston90}.  In order to facilitate a comparison of
results, however, we have decided to use the same value $\beta_c=5.09$ for both
$N_z=16$ and 40 lattices with $N_t=2$. For $N_t=4$ we use the
estimate $\beta_c=5.6924$ obtained on a $24^2\times36\times4$ 
lattice\cite{qcdpax91}.

\begin{figure}[p]
%\vspace{10cm}
\epsfxsize=12cm \epsfbox{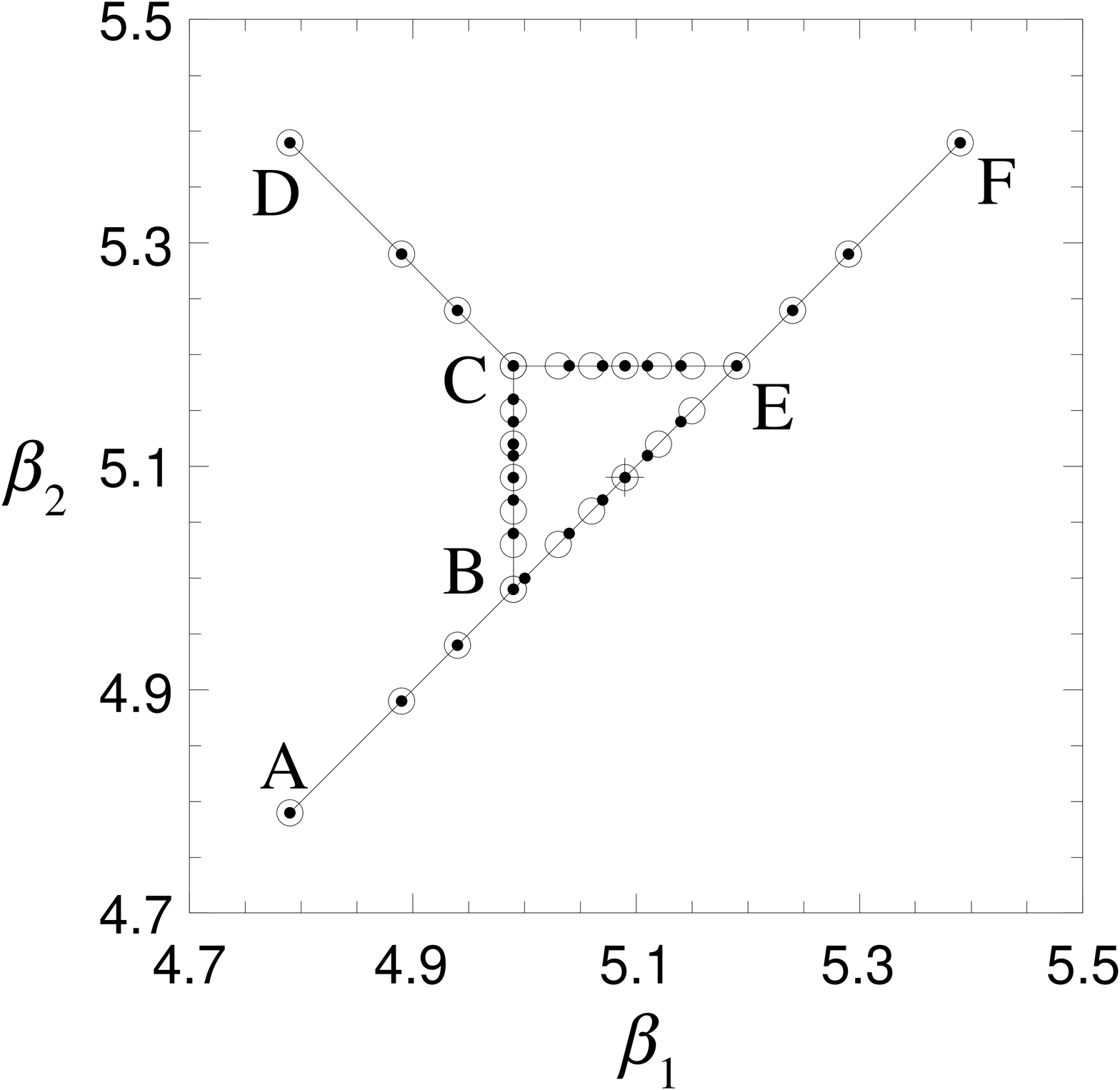}
\caption{{
The Monte Carlo simulation points for $N_t=2$ in the space 
of $(\beta_{+},\beta_{-})$.
Small filled circles are the simulation points for 
an $8^2 \times 16 \times 2$ lattice, and open circles for an 
$8^2 \times 40 \times 2$ lattice. 
The symbol ``+'' corresponds to the phase transition point 
$(\beta_c,\beta_c)$.
}}
\protect\label{fig:intpath}
\end{figure}

In Table ~\ref{tab:param-int16} -- \ref{tab:param-int40} we list the
parameters of our runs on $N_t=2$ lattices together with the
plaquette expectation values obtained.  In Fig.~\ref{fig:intpath}
the corresponding points in the $\beta_1-\beta_2$ plane are plotted. In these
tables  the first group of runs with $\beta_1=\beta_2$ corresponds to the 
simulation points on the line AF of Fig.~\ref{fig:intpath}. 
The second group of runs located on the line CD are used for measuring the
interface tension with the operator method.  
For the integral method we  follow
ref.~\cite{Boston90} and modify the integration path as shown in Fig.~\ref{fig:intpath}, namely
the path ABCD is used for the first integration in (\ref{eq:int-path}) and 
DCEF for the second one. The results along the paths BC and CE are collected in
the third and fourth group in  
Tables~\ref{tab:param-int16} and \ref{tab:param-int40}.  With these runs we can
measure the interface tension for four values of  $\delta\beta=0.1$, 0.15, 0.2,
and 0.3.   The runs on an $N_t=4$ lattice are summarized in
Table~\ref{tab:param-intT4}.  They cover the cases  $\delta\beta=0.02$,
0.03, 0.04, and 0.06. The integration contour is similar to those of Fig.~\ref{fig:intpath}. 
The runs for $N_t=2$ were made on the 16 processor version of QCDPAX and those
for $N_t=4$ on QCDPAX itself\cite{QCDPAX}.

For error estimation of the interface tension in the operator
method we apply the jackknife procedure.  An analysis of the bin 
size dependence
showed that the estimated error is  quite stable
over a wide  range of bin size of  O(100) -- O(10 000) sweeps.  
The errors we quote are for the bin size of about 1000 sweeps.  
Near the critical point, larger bin size is chosen.
Errors of plaquette expectation
values in Table~\ref{tab:param-int16}-- \ref{tab:param-intT4} are also
calculated with the same bin size. 

In the integral method we use the first-order formula to numerically
integrate the plaquette averages.  Jackknife errors of plaquette
averages are combined quadratically  to obtain the statistical
error of the integral.   
We should note that a finite integration step size 
causes a systematic error in addition to the statistical one.  We estimate this
error by comparing the results of the first-order integration with those of the
natural spline fit of the integrand. 
In the worst case, the systematic error estimated in this manner is 
of comparable magnitude with the statistical error 
computed for the first-order formula; 
in most cases, however, 
the systematic error is much smaller than the statistical error.
For the final values of errors, we add the systematic 
error to the statistical one.

%%%%%%%%%%%%%%%%%%%%%%%%%%%%%%%%%%%%%%%%%%%%%%%%%%%%%%%%%%%%%%%%
\section{Results for $N_t=2$}       \label{Results2}

%%%%%%%%%%%%%%%%%%%%%%%%%%%%%%%%%%%%%%%%%%%%%%%%%%%%%%%%%%%%%%%%
\subsection{Operator method}
In Table~\ref{tab:sigma-op} we list the results of the operator method for
$\sigma/T_c^3 $ for
$N_t=2$ lattices.  The one-loop values (\ref{eq:asymmetry}) are used for the
response functions.  Our estimate of $\sigma/T_c^3$ at $\delta\beta=0$ obtained
by a linear extrapolation in  $\delta\beta$ is also given.  

\begin{figure}[p]
\epsfxsize=12cm \epsfbox{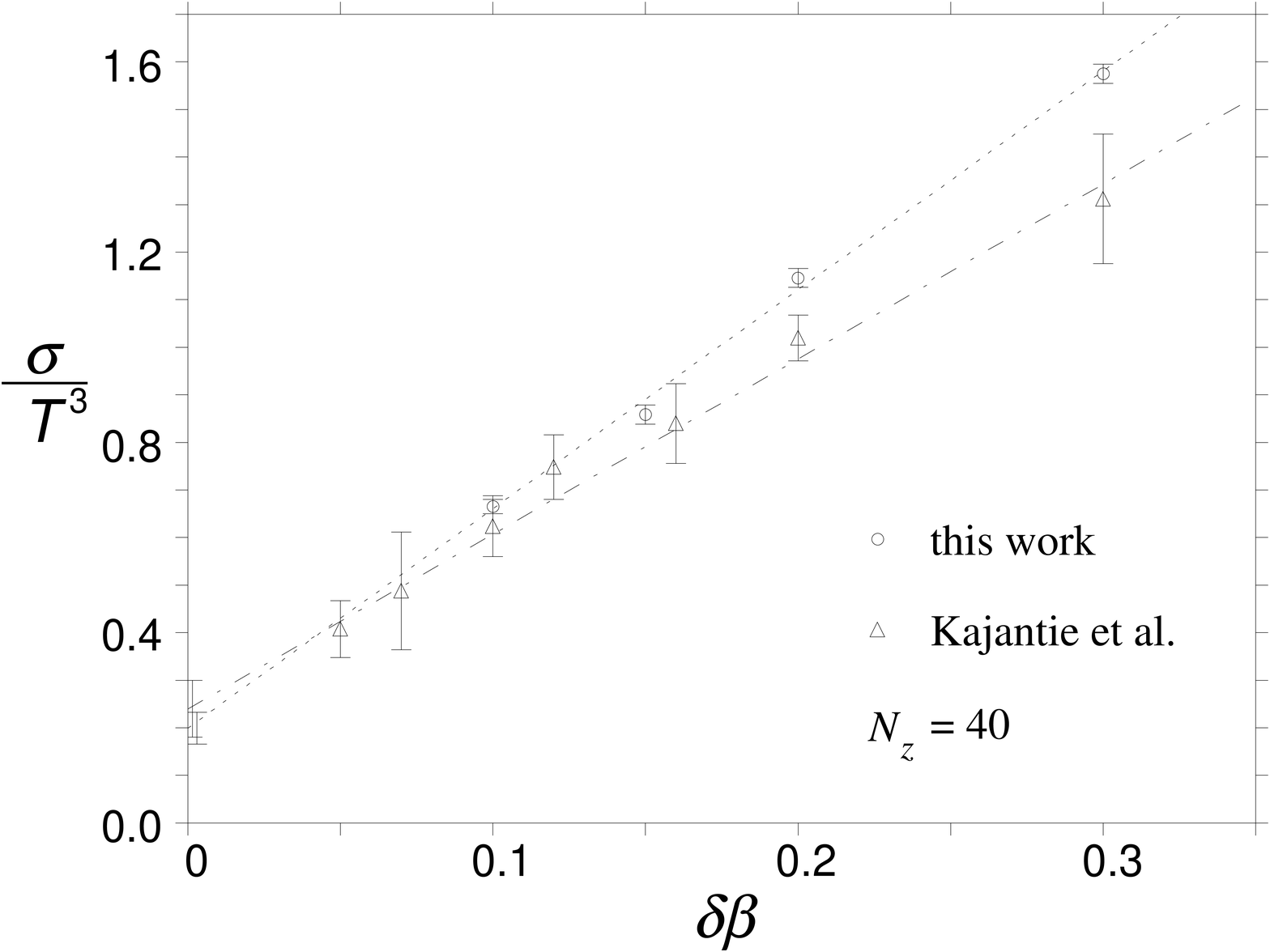}
\caption{{
$\sigma$ with the operator method as a function of $\delta\beta$ 
on an $8^2 \times 40 \times 2$ lattice. 
Open circles are for our results and triangles are for those 
of Kajantie {\it et al.}\protect\cite{Helsinki90}. %[3]. %\cite{Helsinki90}
}}
\protect\label{fig:sigma-op40}
\end{figure}

In Fig.~\ref{fig:sigma-op40} we compare our data on an $8^2\times 40\times 2$ lattice with
those of Kajantie {\it et al.}\cite{Helsinki90} on the same size of lattice.  
We observe that our results at a finite $\delta\beta$ are larger than their
values.  The discrepancy is ascribed  to the difference in the  assignment of
the coupling constant for creating the interface as was
discussed in Sect.~\ref{method}.  The discrepancy diminishes for a
smaller $\delta\beta$, and our extrapolated value
$\sigma/T_c^3=0.199(34)$ is consistent with the result 0.24(6) by
Kajantie {\it et al}.  

%% size dependence %%

\begin{figure}[p]
\epsfxsize=12cm \epsfbox{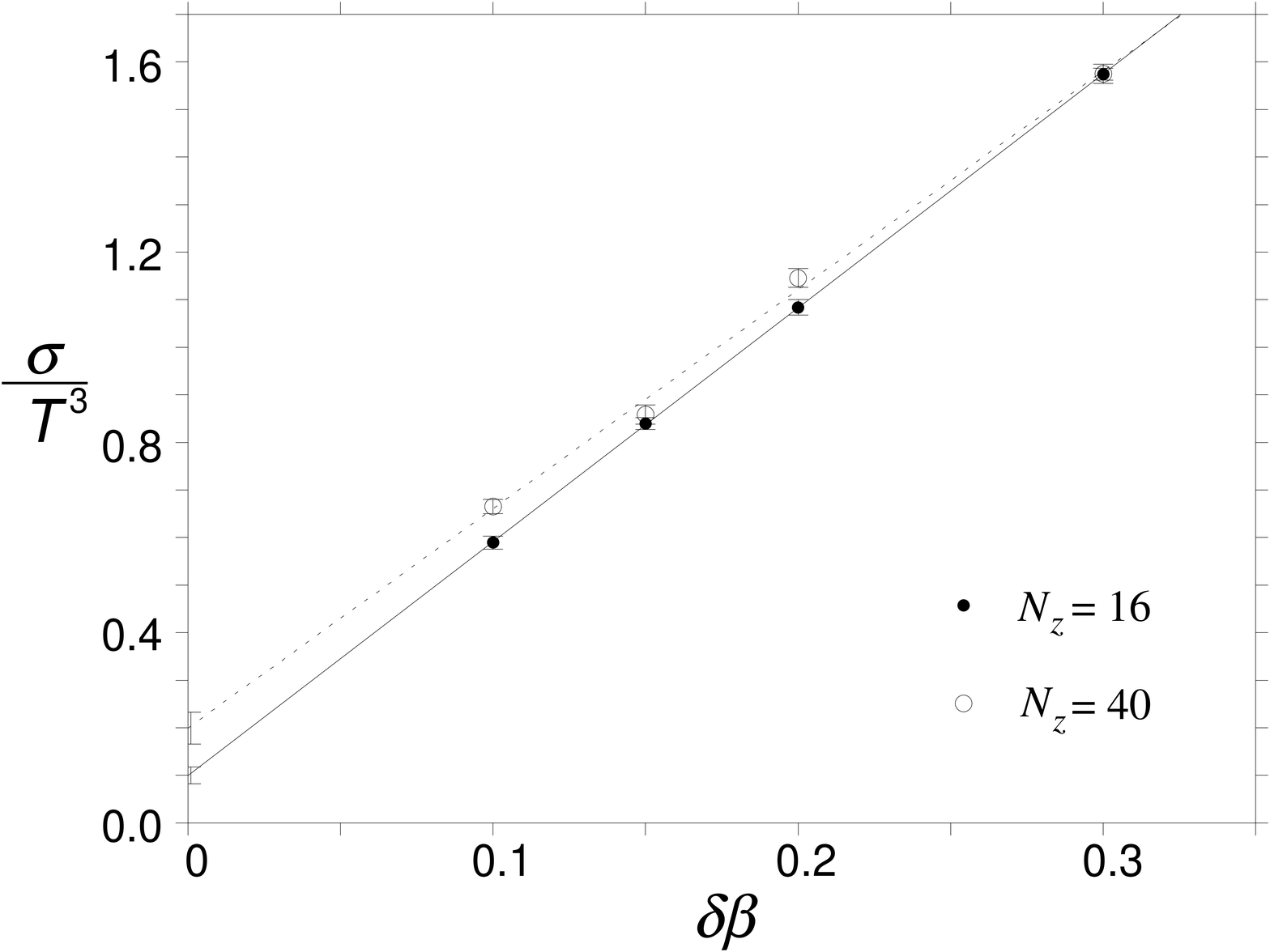}
\caption{{
$\sigma$ with the operator method as a function of $\delta\beta$. 
Small filled circles are the results for the $8^2 \times 16 \times 2$ 
lattice, and open circles for the $8^2 \times 40 \times 2$ lattice. 
}}
\protect\label{fig:sigma-op}
\end{figure}

Our results on an $8^2\times 16\times 2$ lattice is smaller than those on an 
$8^2\times 40\times 2$ lattice as shown in Fig~\ref{fig:sigma-op}.  
The difference increases with
decreasing $\delta\beta$, and a linear extrapolation toward $\delta\beta=0$
results in a factor two difference in the values of $\sigma/T_c^3$ for the two
lattice sizes (see the last column of Table~\ref{tab:sigma-op}).  

To understand the origin of this size dependence, we study the 
spatial thickness of the interface. 
In Fig.~\ref{fig:polyakov-z} we show the $z$-dependence of the 
Polyakov loop rotated to the nearest $Z(3)$ axis $\omrot(N_z)$ for various $\delta\beta$. 
For the larger lattice with $N_z=40$ we observe a clear plateau away from the 
the interface centers  at $z=0$ and $N_z/2$ for all values of $\delta\beta$. 
In contrast, a plateau is barely visible for the $N_z=16$ lattice even
for $\delta\beta=0.3$.   Furthermore, for $\delta\beta
\simm{<} 0.15$, 
$\omrot(16)$ 
deviates not only from the asymptotic values of $\omrot(40)$ 
but also from $\omrot(40)$ at the same distance from the 
interface center even at the middle points $z=4$ and 12 between interfaces.
These observations suggest that 
the half width of the interface region is larger than 4 
and consequently the two interfaces are affecting each other on the $N_z=16$
lattice. 

Similar behavior is also found for the $z$-dependence of the interface free
energy, albeit with larger errors.  In 
Fig.~\ref{fig:sigma-op-width} we plot the quantity $\tilde\sigma(d)$ obtained 
by restricting the  summation in (\ref{eq:operator}) for the interface tension 
to the two regions within a distance  $d$ from the interface centers at $z=0$
and $N_z/2$.  
For $\delta\beta=0.1$, 
we find that $\tilde\sigma(d)$ on the $N_z=40$ lattice saturates  only after 
  $d=5-6$, 
and that $\tilde\sigma(d)$ for $N_z=16$ is clearly not saturated within the
distance $d\leq 4$ available on this lattice.  Furthermore the value of 
$\tilde\sigma(d)$ for $N_z=16$ is smaller than that  for $N_z=40$ at $d
\sim 2$ -- 4.  
From these observations, we conclude that 
$N_z=16$ is not large enough to have two free interfaces within 
the spatial extent of the lattice. We also note that the smaller interface free
energy observed for $N_z=16$ corresponds to an attractive interaction  between
opposite interfaces.

%% non-perturbative Ct and Cs %%

Another problem with the results in Table~\ref{tab:sigma-op} is that they were
obtained employing one-loop perturbative results for the response functions
$c_t$ and $c_s$ and the lattice bare coupling $\beta$ in (\ref{eq:operator}).
This represents a questionable point 
since the critical coupling constant on an $N_t=2$ lattice is large.  
In order to  examine how non-perturbative effects possibly affect the values of
response functions\cite{anisotropy}, we recall that the
combination $c_t+c_s$ is proportional to the scaling beta function: 
$c_t+c_s = - {a \over 2} ({ {\partial \beta} \over {\partial a}} ).$ 
A non-perturbative estimate of this quantity is available from Monte-Carlo
renormalization group studies\cite{mcrg}, which  shows a large deviation of the
$\beta$ function from the perturbative result for $\beta \simm{<} 6$.  % 
The coefficients $c_t$ and $c_s$ also appear in 
other thermodynamic quantities,  the energy density $\epsilon$ 
and the pressure $p$\cite{bielefeld82}. 
In fact the combination $\epsilon - 3p$ is proportional to $c_t+c_s$, while 
$\epsilon + p$ depends both on $c_t+c_s$ and $c_t-c_s$. 
The problem with perturbative estimates of response functions manifests here not
only in a scaling violation of $\epsilon$  and $p$ but also in a
non-vanishing pressure gap at the first order deconfining transition 
\cite{bielefeld90}. 
We can therefore determine $c_t-c_s$ at $\beta_c$ by combining the results for
the beta function from Monte-Carlo renormalization group studies with the
requirement of a vanishing pressure gap.  

Applying this procedure to the data obtained on $24^2 \times 36 \times 4$
and  $36^2 \times 48 \times 6$ lattices \cite{qcdpax91}, 
we find that 
\begin{eqnarray}
c_t-c_s & = & -0.52(12)\qquad\beta=5.6925\nonumber\\
        & = & -0.37(16)\qquad\beta=5.8936
\label{eq:nonpert}
\end{eqnarray} 
($c_t=-0.24(6)$ and $-0.16(8)$ and $c_s=0.28(6)$ and $0.21(8)$ for the two
values of $\beta$, respectively.) These values should be compared with the
perturbative value,  \begin{equation}
c_t-c_s=-0.3336 + O(\beta^{-1}).
\label{eq:pert}
\end{equation}
The non-perturbative estimates (\ref{eq:nonpert}) imply that  the value for
the  interface tension will be  reduced  by a factor  of
0.88(8) for $N_t=4$ and 0.98(10) for$N_t=6$ from that with the perturbative 
value using (\ref{eq:pert}).   This leads us to expect a
 reduction of the interface tension  by  perhaps an even larger magnitude for
$N_t=2$ if non-perturbatively determined response functions are employed.   A
quantitative estimation of the magnitude of reduction, however, requires the
value of the effective beta function at  the critical coupling $\beta_c=5.09$
which is not available at present.

We should note that the  discussion and the conclusion from
Fig.~\ref{fig:sigma-op-width} concerning finite-size effects are not
affected by the correction to $c_t$ and $c_s$ because the main effect 
of it is a shift of the overall coefficient of $\sigma$.

\subsection{Integral method} 

In Table~\ref{tab:sigma-int} we summarize our results for
$\sigma/T_c^3$ for the integral method.  As is seen in Fig.~\ref{fig:sigma-int}
the data lie on a straight line with respect to $\delta\beta$.  The results
of a linear extrapolation are also given in Table~\ref{tab:sigma-int}.  In
Figs.~\ref{fig:plaq-int16} and \ref{fig:plaq-int40} we plot the plaquette
expectation values in the two halves of lattice as a function of $\beta$ along
the integration path for the case of $\delta\beta=0.1$.  The area surrounded by
the two lines gives an estimate of $\sigma$.  

As in the case of the operator method we find that our values for $N_z=16$ has a
different slope from  the corresponding results of Huang {\it et
al.}\cite{Boston90}  at finite $\delta\beta$ (compare filled circles and
triangles in Fig.~\ref{fig:sigma-int}), while the value extrapolated to
$\delta\beta=0$  is perfectly consistent, namely we find $\sigma/T_c^3=0.113(6)$
as compared to $\sigma/T_c^3=0.115(13)$ in ref.~\cite{Boston90}.

The results on an $8^2\times 40\times 2$ lattice are larger by about 5-10\%
from those on an $8^2\times 16\times 2$.  The difference increases to
20\% after extrapolation to $\delta\beta=0$, which, however, is much
smaller compared to the case of the operator method where a factor two
discrepancy is observed.  
We also note that the size effect is not sensitive to the value of 
$\delta\beta$ (Fig.~\ref{fig:sigma-int}), 
in contrast to the case of the operator method discussed in the previous 
subsection. 

The origin of this difference of lattice size dependence may be 
understood in the following way.
We expect the size effect on the interface to increase 
when both $\beta_1$ and $\beta_2$ 
become  close to $\beta_c$ while keeping $\beta_1 < \beta_c < \beta_2$. 
%
%For the combinations $(\beta_1,\beta_2)$ used in the integral method, 
With our choice of the integration path as shown in 
Fig.~\ref{fig:intpath}, the size effect is therefore largest at the point C,
diminishing toward the points D, B and E.  Near the points B and E and along
the paths BA and EF  we expect size effects to be small because of the absence of
the interface. 
We now recall that the interface tension in the integral method is obtained
from the integrals along the paths ABCD and DCEF, while that with the operator
method is taken at a point along the path DC.  We then expect the magnitude of
size effects to be smaller in the integral method since the large size effect
near the point C, directly affecting results for the operator method,  is
diluted by contribution from other parts of the paths suffering less from the
effects.   Furthermore the fact that  the paths BC and CE are common to all of
our choices of $\delta\beta$ explains a similar magnitude of size effects
seen in Fig.~\ref{fig:sigma-int}.

Comparing the results for $N_z=40$,
we find that the value of $\sigma/T_c^3$ obtained by the operator method 
(with the perturbative coefficients) is larger than that of the integral 
method by a factor of about 1.5. 
We ascribe this discrepancy to the presence of non-perturbative corrections to
the response functions based on the consideration discussed in
the previous subsection. 

We also note that our result $\sigma/T_c^3=0.134(16)$ with
the integral method  is consistent with a value $\sigma/T_c^3=0.139(4)$ from
a determination with a transfer matrix method  on $4^4 \times
64 \times 2$ --  $8^2 \times 128 \times 2$ lattices\cite{Grossmann92b}, 
while a determination with a histogram  method  on 
$8^2 \times 30 \times 2$ -- $14^2 \times 42 \times 2$ lattices 
\cite{Grossmann93} gives a slightly smaller value 0.092(4).

\section{Results for $N_t=4$}	\label{Results4}

Our comparative analysis of the operator and integral methods for $N_t=2$
lattices shows that the integral method suffers much less from  various
systematic errors.  We therefore adopt the integral method for our study on an
$N_t=4$ lattice.

A difficulty in an extraction of the interface tension for $N_t=4$ is the
smallness of the value of $\sigma a^3$.  In fact assuming
scaling the result  $\sigma/T_c^3\sim 0.1$ for $N_t=2$ leads to $\sigma
a^3\sim 0.0015$ for $N_t=4$. The actual value appears even smaller;
the value  $ \sigma/T_c^3 = 0.027(4) $ \cite{Brower92} obtained with the 
integral method using an external field coupled to the Polyakov loop 
translates to $\sigma a^3=0.00042(6)$.  Thus  very accurate results of plaquette
expectation values for $\delta\beta$ much smaller than those employed for the
case of $N_t=2$ will be needed. Furthermore it is known that  a spatial size of
at least 4--5 times larger than the temporal size is needed for clear first-order
signals on an $N_t=4$ lattice\cite{puresu3,qcdpax91}.  
This requirement was not  met in
the first attempt \cite{Boston90} toward an interface tension measurement on an $N_t=4$ lattice,
which  failed to find a finite value of $\sigma$ employing $8^2 \times 16 \times
4$ and $12^2 \times 24 \times 4$ lattices.    
Based on those
considerations we chose to work on an \NtFourLattice\ lattice and carried out
simulations for $\delta\beta=0.02$ -- 0.06. 

As discussed in the previous section, a large $N_z$ is required 
to remove the effect of interface thickness. 
In Fig.~\ref{fig:poly-T4} we plot  $\omrot$ for 
$\delta\beta=0.02$ and 0.04 as a function of $z$. 
We observe a plateau for both cases, albeit somewhat limited in range for 
$\delta\beta=0.02$.  
From our experience with the $N_t=2$ lattices we consider that 
the
effect of  interface thickness is sufficiently suppressed for 
$\delta\beta \simm{>} 0.02$

Our result for $\sigma$ is summarized in Table~\ref{tab:sigma-intT4} and
plotted in Fig.~\ref{fig:sigma-intT4}. Taking the three points with
$\delta\beta\leq 0.04$ and extrapolating linearly toward $\delta\beta=0$ we
find $\sigma/T_c^3=0.0171(113)$.  The central  value is smaller than the previous
estimate $ \sigma/T_c^3 = 0.027(4) $ obtained  on  a $16^2 \times 32 \times 4$
lattice \cite{Brower92}.  We note that a recent study with the  histogram method
using the data of \cite{qcdpax91} on lattices  $12^2 \times 24 \times 4$ and
$24^2 \times 36 \times 4$  also gives a larger value
0.0292(22) \cite{CERN-QCDPAX93}.  
The error in our result, however, is too large
to see if there is a real discrepancy.

\section{Conclusions}	\label{Conclusion}

We have studied the confinement-deconfinement interface tension in  
the pure $SU(3)$ gauge theory on an $8^2\times N_z\times 2$
lattice with $N_z=16$ and 40 by applying the operator and 
integral methods.  Our results confirm the values reported previously when
compared at the same parameters.
Our study of the $N_z$ dependence shows, however,  that the thickness
of the interface causes  misleadingly smaller values for the interface tension
$\sigma$ when estimated on a lattice of a small length such as $N_z=16$.  This
effect is severer for the operator method, which  explains 
part of the discrepancy in the interface tension obtained with the 
two methods. 
We made some quantitative estimates on non-perturbative corrections 
to the responce function $c_t-c_s$ needed in the operator method, 
and argued that incorporating them would remove the rest of the
discrepancy.
We conclude that the integral method yields more reliable values for the
interface tension, suffering less from various systematic errors.  
Our result $\sigma/T_c^3=0.134(16)$ with the integral method on a 
large lattice is consistent with a determinations with a transfer matrix 
method\cite{Grossmann92b}. 

We have also attempted a measurement of the interface tension on an 
$N_t=4$ lattice with the integral method.  The result is smaller than
those reported in the literature\cite{Brower92,CERN-QCDPAX93}.  It has
quite a large error, however, 
in spite of the substantial statistics accumulated
with 14 days of fully  running QCDPAX.  The problem originates from a rapid
decrease of the discontinuity in plaquette averages with an increasing
$N_t$.  A judicious choice of observables having a larger jump across the
deconfining transition as was done by Brower {\it et al.}\cite{Brower92} will
be  indispensable for a determination of the interface tension closer to the
limit of continuum space-time.

\section*{Acknowledgements} % 
We are grateful to A.\ Ukawa for valuable discussions and suggestions and 
continuous encouragement.
We also thank the members of the QCDPAX collaboration for 
discussions and their help. 
This project is in part supported by the Grant-in-Aid
of Ministry of Education, Science and Culture
(No.62060001 and No.02402003) 
and by the University of Tsukuba Research Projects.

%%%%%%%%%%% references %%%%%%%%%%%%%%%%%%%%%%%%%%%%%%%%%%%

%%%%%%%%%%%%%%%%%%%%%%%%%%%%% table %%%%%%%%%%%%%%%%%%%%%%%%%%%

%%%%%%%%%%%%%%%%%%%%%%%%%%%%% table %%%%%%%%%%%%%%%%%%%%%%%%%%%
\newpage

\begin{table}[p]
% \begin{center}
\begin{tabular}
{c@{\hspace{2.5mm}}c|rr|l@{\,(\,}r@{)\hspace{2.5mm}}l@{\,(\,}r@{)\hspace{2.5mm}}l@{\,(\,}r@{)}}
   \hline   \hline
     & & \multicolumn{2}{c|} {run parameters} & \multicolumn{6}{c}{} \\
   \cline{3-4}
        {\raisebox{1.5ex}[0pt] {$\beta_1$} }
      & {\raisebox{1.5ex}[0pt] {$\beta_2$} }
      & \multicolumn{1}{r}{tot.\ sw.}
      & \multicolumn{1}{r|}{therm.}
      & \multicolumn{2}{c}{\raisebox{1.5ex}[0pt]{$P_1$}}
      & \multicolumn{2}{c}{\raisebox{1.5ex}[0pt]{$P_2$}}
      & \multicolumn{2}{c}{\raisebox{1.5ex}[0pt]{$P_2 - P_1$}}
   \\ \hline
   \multicolumn{2}{c|}{4.79} & 21000 & 1000 & \multicolumn{4}{c}{0.373370 (45)} \\
   \multicolumn{2}{c|}{4.89} & 21000 & 1000 & \multicolumn{4}{c}{0.386170 (37)} \\
   \multicolumn{2}{c|}{4.94} & 21000 & 1000 & \multicolumn{4}{c}{0.392992 (41)} \\
   \multicolumn{2}{c|}{4.99} & 30000 & 1000 & \multicolumn{4}{c}{0.400056 (46)} \\
   \multicolumn{2}{c|}{5.00} & 30000 & 1000 & \multicolumn{4}{c}{0.401590 (48)} \\
   \multicolumn{2}{c|}{5.04} & 30000 & 1000 & \multicolumn{4}{c}{0.407723 (55)} \\
   \multicolumn{2}{c|}{5.07} & 30000 & 1000 & \multicolumn{4}{c}{0.41261\hspace{3.5mm}(10)} \\
   \multicolumn{2}{c|}{5.09} & 198000 & 8000 & \multicolumn{4}{c}{0.4278\hspace{5.5mm}(15)} \\
   \multicolumn{2}{c|}{5.11} & 30000 & 1000 & \multicolumn{4}{c}{0.45551\hspace{3.5mm}(22)} \\
   \multicolumn{2}{c|}{5.14} & 30000 & 1000 & \multicolumn{4}{c}{0.46652\hspace{3.5mm}(12)} \\
   \multicolumn{2}{c|}{5.19} & 30000 & 1000 & \multicolumn{4}{c}{0.480807 (95)} \\
   \multicolumn{2}{c|}{5.24} & 30000 & 1000 & \multicolumn{4}{c}{0.492618 (75)} \\
   \multicolumn{2}{c|}{5.29} & 21000 & 1000 & \multicolumn{4}{c}{0.503312 (75)} \\
   \multicolumn{2}{c|}{5.39} & 24000 & 4000 & \multicolumn{4}{c}{0.521825 (68)} \\
   \hline
   4.99 & 5.19 & 200000 & 10000 & 0.407549 & 86 & 0.470238 & 92 & 0.062689 & 79 \\
   4.94 & 5.24 & 200000 & 10000 & 0.399601 & 45 & 0.482138 & 60 & 0.082537 & 52 \\
   4.89 & 5.29 & 125000 & 10000 & 0.392557 & 42 & 0.492648 & 59 & 0.100091 & 56 \\
   4.79 & 5.39 & 200000 & 10000 & 0.379668 & 27 & 0.511044 & 35 & 0.131376 & 37 \\
   \hline
   4.99 & 5.04 & 30000 & 3000 & 0.400363 & 64 & 0.407368 & 60 \\
   4.99 & 5.07 & 60000 & 3000 & 0.400476 & 50 & 0.411974 & 58 \\
   4.99 & 5.09 & 60000 & 3000 & 0.400610 & 45 & 0.415267 & 67 \\
   4.99 & 5.11 & 60000 & 3000 & 0.400846 & 61 & 0.41970 & 31 \\
   4.99 & 5.12 & 60000 & 3000 & 0.40124 & 10 & 0.42504 & 77 \\
   4.99 & 5.14 & 60000 & 3000 & 0.40338 & 15 & 0.4459 & 10 \\
   4.99 & 5.16 & 60000 & 3000 & 0.40539 & 14 & 0.45930 & 35 \\
     \hline
   5.04 & 5.19 & 60000 & 3000 & 0.42676 & 32 & 0.47455 & 15 \\
   5.07 & 5.19 & 60000 & 3000 & 0.44340 & 31 & 0.47709 & 11 \\
   5.09 & 5.19 & 60000 & 3000 & 0.4519 & 16 & 0.477845 & 97 \\
   5.11 & 5.19 & 60000 & 3000 & 0.45917 & 12 & 0.478778 & 82 \\
   5.14 & 5.19 & 30000 & 3000 & 0.46769 & 14 & 0.47928 & 11 \\
   \hline
   \hline
\end{tabular}
%\end{center}
\caption{Run parameters and plaquette expectation values in 
each domain on the {$ 8^2 \times 16 \times 2$} lattice. 
These simulation points are plotted in Fig.~1. 
\protect\label{tab:param-int16}
}
%\medskip\noindent
\end{table}

%%%%%%%%%%%%%%%%%%%%%%%%%%%%% table %%%%%%%%%%%%%%%%%%%%%%%%%%%

\begin{table}[p]
%\begin{center}
\begin{tabular}
{c@{\hspace{2.5mm}}c|rr|l@{\,(\,}r@{)\hspace{2.5mm}}l@{\,(\,}r@{)\hspace{2.5mm}}l@{\,(\,}r@{)}}
   \hline
   \hline
     & & \multicolumn{2}{c|} {run parameters} & \multicolumn{6}{c}{} \\
   \cline{3-4}
        {\raisebox{1.5ex}[0pt] {$\beta_1$} }
      & {\raisebox{1.5ex}[0pt] {$\beta_2$} }
      & \multicolumn{1}{r}{tot.\ sw.}
      & \multicolumn{1}{r|}{therm.}
      & \multicolumn{2}{c}{\raisebox{1.5ex}[0pt]{$P_1$}}
      & \multicolumn{2}{c}{\raisebox{1.5ex}[0pt]{$P_2$}}
      & \multicolumn{2}{c}{\raisebox{1.5ex}[0pt]{$P_2 - P_1$}}
   \\ \hline 
   \multicolumn{2}{c|}{4.79} & 22000 & 4000 & \multicolumn{4}{c}{0.373399 (23)} \\
   \multicolumn{2}{c|}{4.89} & 20000 & 2000 & \multicolumn{4}{c}{0.386199 (28)} \\
   \multicolumn{2}{c|}{4.94} & 20000 & 2000 & \multicolumn{4}{c}{0.392953 (31)} \\
   \multicolumn{2}{c|}{4.99} & 20000 & 2000 & \multicolumn{4}{c}{0.400122 (35)} \\
   \multicolumn{2}{c|}{5.03} & 20000 & 2000 & \multicolumn{4}{c}{0.406027 (45)} \\
   \multicolumn{2}{c|}{5.06} & 20000 & 2000 & \multicolumn{4}{c}{0.410851 (48)} \\
   \multicolumn{2}{c|}{5.09} & 100000 & 10000 & \multicolumn{4}{c}{0.41918\hspace{3.5mm}(99)} \\
   \multicolumn{2}{c|}{5.0937} & 530000 & 30000 & \multicolumn{4}{c}{0.4306\hspace{5.5mm}(18)} \\
   \multicolumn{2}{c|}{5.12} & 20000 & 2000 & \multicolumn{4}{c}{0.45931\hspace{3.5mm}(15)} \\
   \multicolumn{2}{c|}{5.15} & 20000 & 2000 & \multicolumn{4}{c}{0.46963\hspace{3.5mm}(10)} \\
   \multicolumn{2}{c|}{5.19} & 20000 & 2000 & \multicolumn{4}{c}{0.480845 (80)} \\
   \multicolumn{2}{c|}{5.24} & 20000 & 2000 & \multicolumn{4}{c}{0.492589 (62)} \\
   \multicolumn{2}{c|}{5.29} & 20000 & 2000 & \multicolumn{4}{c}{0.503317 (47)} \\
   \multicolumn{2}{c|}{5.39} & 22000 & 4000 & \multicolumn{4}{c}{0.521852 (41)} \\
   \hline
   4.99 & 5.19 & 400000 & 20000 & 0.402641 & 20 & 0.476307 & 29 & 0.073666 & 26 \\
   4.94 & 5.24 & 200000 & 10000 & 0.395490 & 19 & 0.488416 & 29 & 0.092926 & 30 \\
   4.89 & 5.29 & 200000 & 10000 & 0.388714 & 15 & 0.499055 & 25 & 0.110341 & 26 \\
   4.79 & 5.39 & 200000 & 10000 & 0.375886 & 13 & 0.517555 & 20 & 0.141669 & 23 \\
   \hline
   4.99 & 5.03 & 26000 & 5000 & 0.400123 & 73 & 0.405956 & 66 \\
   4.99 & 5.06 & 24000 & 3000 & 0.400268 & 43 & 0.410796 & 72 \\
   4.99 & 5.09 & 52000 & 4000 & 0.400280 & 37 & 0.41675 & 39 \\
   4.99 & 5.12 & 24000 & 3000 & 0.400989 & 59 & 0.44999 & 35 \\
   4.99 & 5.15 & 24000 & 3000 & 0.401837 & 58 & 0.46402 & 17 \\
   \hline
   5.03 & 5.19 & 24000 & 3000 & 0.41004 & 12 & 0.47712 & 11 \\
   5.06 & 5.19 & 24000 & 3000 & 0.41804 & 34 & 0.47823 & 19 \\
   5.09 & 5.19 & 74000 & 6000 & 0.44597 & 67 & 0.479589 & 71 \\
   5.12 & 5.19 & 24000 & 3000 & 0.46070 & 15 & 0.480154 & 79 \\
   5.15 & 5.19 & 24000 & 3000 & 0.47006 & 14 & 0.48020 & 10 \\
   \hline
\end{tabular}
%\end{center}
\caption{The same as Table 1 for the {$ 8^2 \times 40 \times 2$} lattice.
}
\protect\label{tab:param-int40}
\medskip\noindent
\end{table}

%%%%%%%%%%%%%%%%%%%%%%%%%%%%% table %%%%%%%%%%%%%%%%%%%%%%%%%%%

\begin{table}[p]
%\begin{center}
\begin{tabular}
{c@{\hspace{2.5mm}}c|rr|l@{\,(\,}r@{)\hspace{2.5mm}}l@{\,(\,}r@{)\hspace{2.5mm}}l@{\,(\,}r@{)}}
   \hline
   \hline
     & & \multicolumn{2}{c|} {run parameters} & \multicolumn{4}{c}{} & \multicolumn{2}{c}{$P_2-P_1$}\\
   \cline{3-4}
        {\raisebox{1.5ex}[0pt] {$\beta_1$} }
      & {\raisebox{1.5ex}[0pt] {$\beta_2$} }
      & \multicolumn{1}{r}{tot.\ sw.}
      & \multicolumn{1}{r|}{therm.}
      & \multicolumn{2}{c}{\raisebox{1.5ex}[0pt]{$P_1$}}
      & \multicolumn{2}{c}{\raisebox{1.5ex}[0pt]{$P_2$}}
      & \multicolumn{2}{c}{$\times 10^2$}
   \\ \hline 
   \multicolumn{2}{c|}{5.6324} & 12000 & 2000 & \multicolumn{4}{c}{0.533130 (38)} \\
   \multicolumn{2}{c|}{5.6524} & 12000 & 2000 & \multicolumn{4}{c}{0.538238 (46)} \\
   \multicolumn{2}{c|}{5.6624} & 12000 & 2000 & \multicolumn{4}{c}{0.540762 (45)} \\
   \multicolumn{2}{c|}{5.6724} & 20000 & 2000 & \multicolumn{4}{c}{0.543113 (32)} \\
   \multicolumn{2}{c|}{5.6804} & 20000 & 2000 & \multicolumn{4}{c}{0.545051 (44)} \\
   \multicolumn{2}{c|}{5.6870} & 20000 & 2000 & \multicolumn{4}{c}{0.54692\hspace{3.5mm}(20)} \\
   \multicolumn{2}{c|}{5.6924} & 80000 & 5000 & \multicolumn{4}{c}{0.54873\hspace{3.5mm}(30)} \\
   \multicolumn{2}{c|}{5.6978} & 20000 & 2000 & \multicolumn{4}{c}{0.552987 (53)} \\
   \multicolumn{2}{c|}{5.7044} & 20000 & 2000 & \multicolumn{4}{c}{0.554618 (51)} \\
   \multicolumn{2}{c|}{5.7124} & 20000 & 2000 & \multicolumn{4}{c}{0.556508 (34)} \\
   \multicolumn{2}{c|}{5.7224} & 12000 & 2000 & \multicolumn{4}{c}{0.558488 (36)} \\
   \multicolumn{2}{c|}{5.7324} & 12000 & 2000 & \multicolumn{4}{c}{0.560475 (32)} \\
   \multicolumn{2}{c|}{5.7524} & 12000 & 2000 & \multicolumn{4}{c}{0.563854 (25)} \\
   \hline
   5.6724 & 5.7124 & 32000 & 4000 & 0.544131 & 67 & 0.555520 & 66 & 1.1390 & 87 \\
   5.6624 & 5.7224 & 32000 & 4000 & 0.541635 & 75 & 0.557411 & 89 & 1.5776 & 57 \\
   5.6524 & 5.7324 & 32000 & 4000 & 0.539321 & 44 & 0.559379 & 35 & 2.0058 & 45 \\
   5.6324 & 5.7524 & 32000 & 4000 & 0.534442 & 33 & 0.562714 & 25 & 2.8272 & 39 \\
   5.6124 & 5.7724 & 32000 & 4000 & 0.529389 & 40 & 0.565739 & 21 & 3.6350 & 47 \\
   \hline
   5.6724 & 5.682 &  60000 & 2000  & 0.543233 &  24 & 0.545364 & 30 \\
   5.6724 & 5.688 &  60000 & 2000  & 0.543296 &  33 & 0.546843 & 63 \\
   5.6724 & 5.694 & 100000 & 4000  & 0.543410 &  52 & 0.54871 & 10 \\
   5.6724 & 5.700 & 120000 & 4000  & 0.543513 &  50 & 0.55092 & 21 \\
   5.6724 & 5.706 &  60000 & 2000  & 0.543827 &  41 & 0.553736 & 77 \\
     \hline
   5.678 & 5.7124 &  60000 & 2000  & 0.545623 &  80 & 0.555678 & 50 \\
   5.684 & 5.7124 & 100000 & 4000  & 0.54831  & 11  & 0.556022 & 29 \\
   5.690 & 5.7124 &  88000 & 4000  & 0.550893 &  83 & 0.556265 & 26 \\
   5.696 & 5.7124 &  40000 & 2000  & 0.552771 &  52 & 0.556342 & 33 \\
   5.702 & 5.7124 &  40000 & 2000  & 0.554249 &  45 & 0.556372 & 37 \\
   \hline
\end{tabular}
%\end{center}
\caption{The same as Table 1 for the {$ 18^2 \times 48 \times 4$} lattice.
The last run in the second group was not used in the integral method. 
}
\protect\label{tab:param-intT4}
\medskip\noindent
\end{table}

%%%%%%%%%%%%%%%%%%%%%%%%%%%%% table %%%%%%%%%%%%%%%%%%%%%%%%%%%

%\smallskip
\begin{table}[p]
\begin{center}
\begin{tabular}{cccccc@{\hspace{0.5mm}}c}    \hline \hline
   &  \multicolumn{4}{c} {$ \delta\beta $} & \\
   \cline{2-5}
   {\raisebox{1.5ex}[0pt] {$N_z$}} 
      & 0.3 & 0.2 & 0.15 & 0.1 &
   {\raisebox{1.5ex}[0pt] {$ \sigma/T_c^3 $}} &
   {\raisebox{1.5ex}[0pt] {$ \chi^2/dof $}} 
   \\ \hline
   16  & 1.575\,(12) & 1.084\,(16) & 0.840\,(13) & 0.590\,(14) & 0.100\,(18) & 0.03 \\
   40  & 1.575\,(20) & 1.146\,(20) & 0.859\,(20) & 0.665\,(15) & 0.199\,(34) & 2.15 \\
   \hline \hline
\end{tabular}
\end{center}
\caption{{$\sigma(\delta\beta)/T_c^3$ for $N_t=2$ 
%%on the $8^2 \times N_z \times 2$ lattice 
at finite $\delta\beta$ determined with the operator method. 
The last two columns are the results of a linear extrapolation 
to $\delta\beta=0$ and its $\chi^2$ per degree of freedom. 
}}
\protect\label{tab:sigma-op}
\medskip\noindent
\end{table}

%%%%%%%%%%%%%%%%%%%%%%%%%%%%% table %%%%%%%%%%%%%%%%%%%%%%%%%%%

%\smallskip
\begin{table}[p]
\begin{center}
\begin{tabular}{cccccc@{\hspace{0.5mm}}c}    \hline \hline
   &  \multicolumn{4}{c} {$ \delta\beta $} & \\
   \cline{2-5}
   {\raisebox{1.5ex}[0pt] {$N_z$}} 
      & 0.3 & 0.2 & 0.15 & 0.1 &
   {\raisebox{1.5ex}[0pt] {$ \sigma/T_c^3 $}} &
   {\raisebox{1.5ex}[0pt] {$ \chi^2/dof $}} 
   \\ \hline
   16  & 1.098\,(4) & 0.771\,(5) & 0.607\,(4) & 0.438\,(7) & 0.113\,(6) & 0.17 \\
   40  & 1.111\,(11) & 0.786\,(13) & 0.624\,(12) & 0.458\,(12) & 0.134\,(16) & 0.01 \\
   \hline \hline
\end{tabular}
\end{center}
\caption{{$\sigma(\delta\beta)/T_c^3$ for $N_t=2$ 
determined with the integral method. 
The last two columns are the results of a linear extrapolation 
to $\delta\beta=0$. 
}}
\protect\label{tab:sigma-int}
\medskip\noindent
\end{table}

%%%%%%%%%%%%%%%%%%%%%%%%%%%%% table %%%%%%%%%%%%%%%%%%%%%%%%%%%

%\smallskip
\begin{table}[p]
\begin{center}
\begin{tabular}{cccccc@{\hspace{0.5mm}}c}  \hline \hline
   \multicolumn{4}{c} {$ \delta\beta $} & & \\
   \cline{1-4}
      0.06 & 0.04 & 0.03 & 0.02 &
   {\raisebox{1.5ex}[0pt] {$ \sigma/T_c^3 $}} &
   {\raisebox{1.5ex}[0pt] {$ \chi^2/dof $}} 
   \\ \hline
   0.6031(122) & 0.3897(67) & 0.2945(68) & 0.2034(45) & 0.0171(113) & 0.06 \\
   \hline \hline
\end{tabular}
\end{center}
\caption{{$\sigma(\delta\beta)/T_c^3$ for $N_t=4$ determined 
with the integral method. 
The last two columns are the results of a linear extrapolation 
to $\delta\beta=0$ using $\delta\beta=0.02$ -- 0.04.
}}
\protect\label{tab:sigma-intT4}
\medskip\noindent
\end{table}

\clearpage

%%%%%%%%%%%%%%%%%%%%%%%%%%%%% fig %%%%%%%%%%%%%%%%%%%%%%%%%%%%%

%%%%%%%%%%%%%%%%%%%%%%%%%%%%% fig %%%%%%%%%%%%%%%%%%%%%%%%%%%%%
\newpage

%%%%%%%%%%%%%%%%%%%%%%%%%%%%%

%%%%%%%%%%%%%%%%%%%%%%%%%%%%%

%%%%%%%%%%%%%%%%%%%%%%%%%%%%%

\begin{figure}[p]
\epsfxsize=12cm \epsfbox{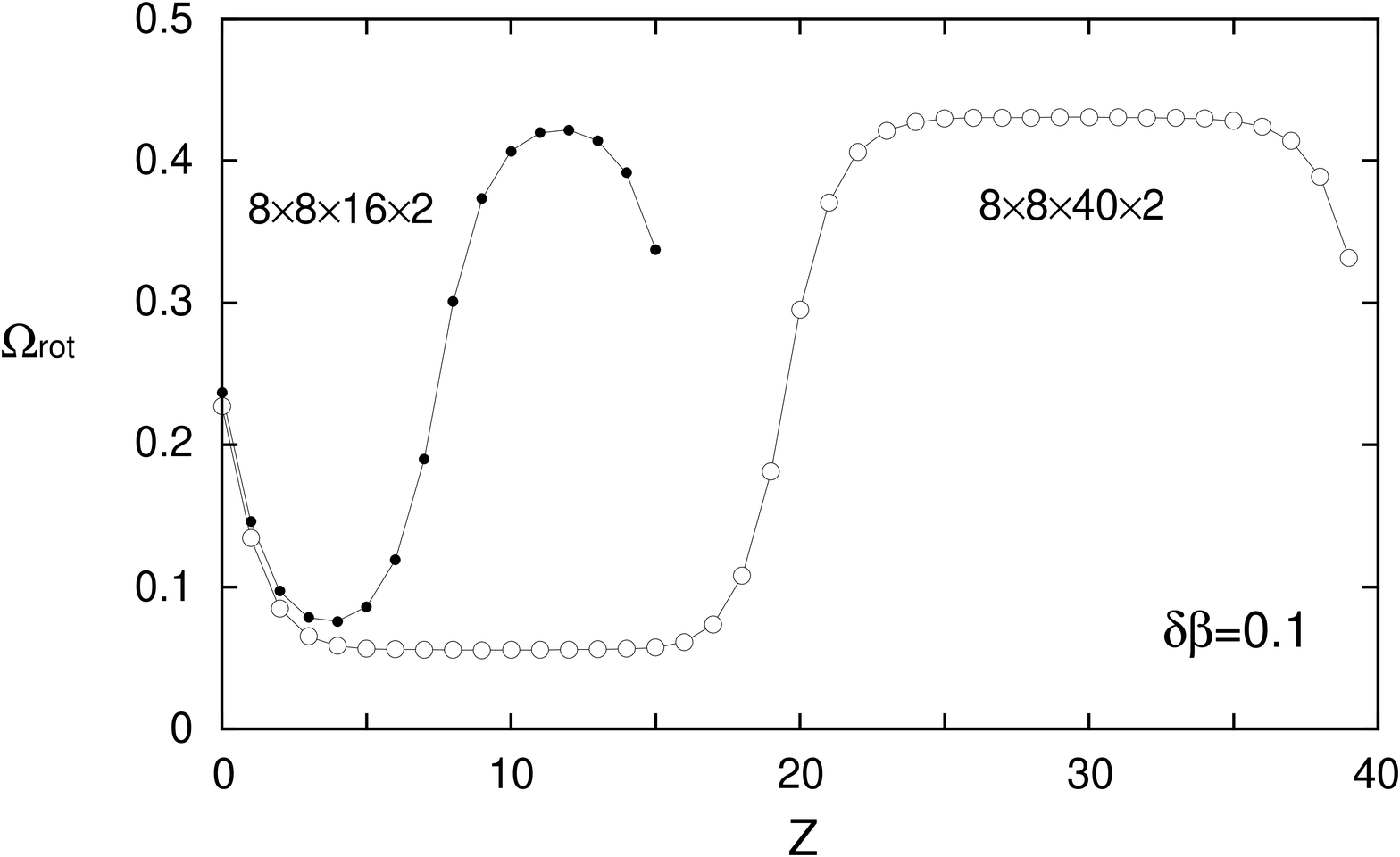}
\epsfxsize=12cm \epsfbox{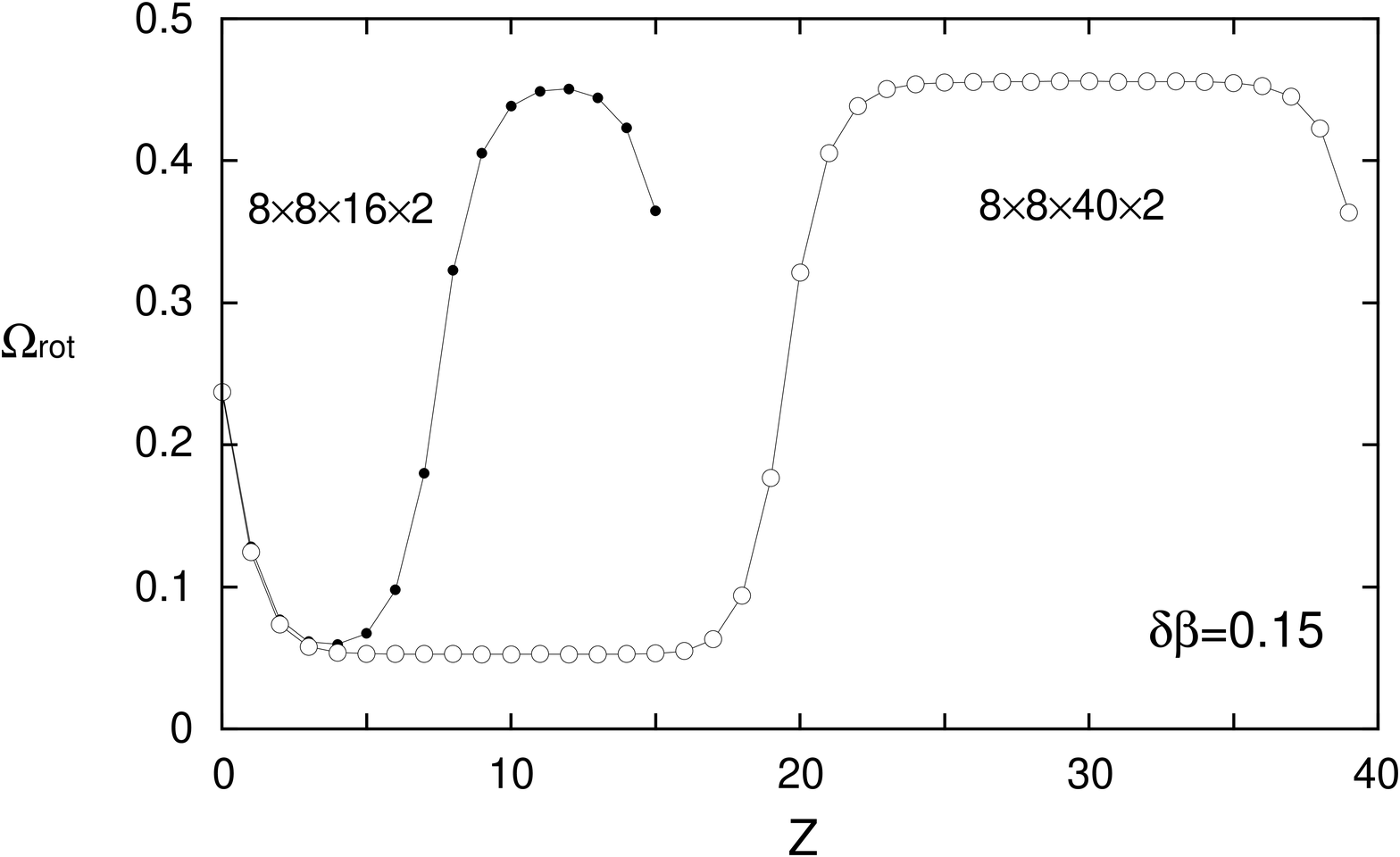}
\caption{{
Real part of the rotated Polyakov loop 
as a function of $z$ 
on the $8^2 \times 16 \times 2$ and $8^2 \times 40 \times 2$ lattices.
The Polyakov loop is rotated by a $Z(3)$ phase factor so that the 
argument is in (-$\pi/3$,$\pi/3$]. 
Errors are much smaller than the symbols. 
(a) for $\delta\beta=0.1$ and (b) $\delta\beta=0.15$.
}}
\protect\label{fig:polyakov-z}
\end{figure}

\addtocounter{figure}{-1}
\begin{figure}[p]
\epsfxsize=12cm \epsfbox{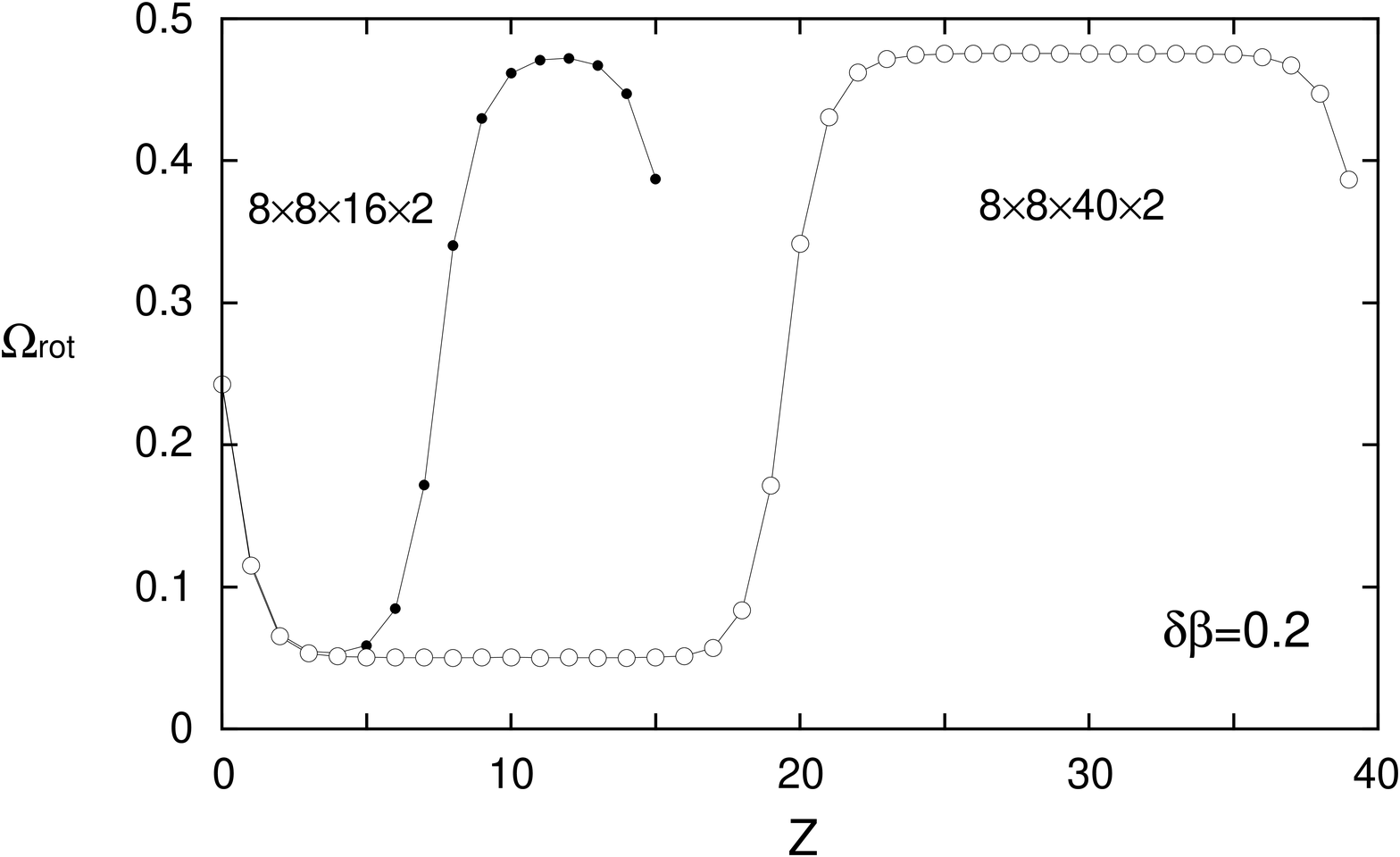}
\epsfxsize=12cm \epsfbox{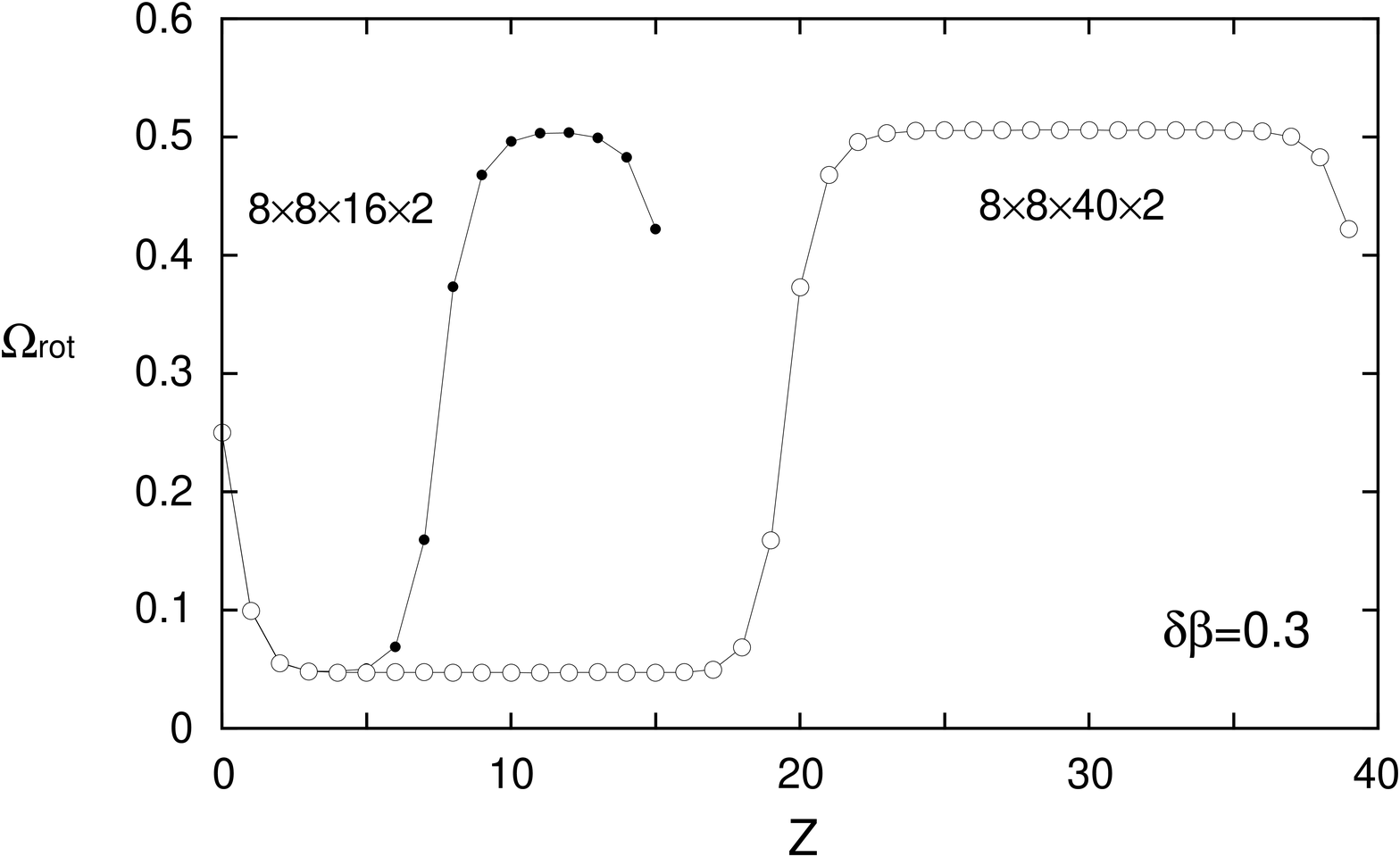}
\caption{{({\it Continued}\/)
(c) $\delta\beta=0.2$ and (d) $\delta\beta=0.3$. 
}}
\end{figure}

%%%%%%%%%%%%%%%%%%%%%%%%%%%%%

\begin{figure}[p]
\epsfxsize=12cm \epsfbox{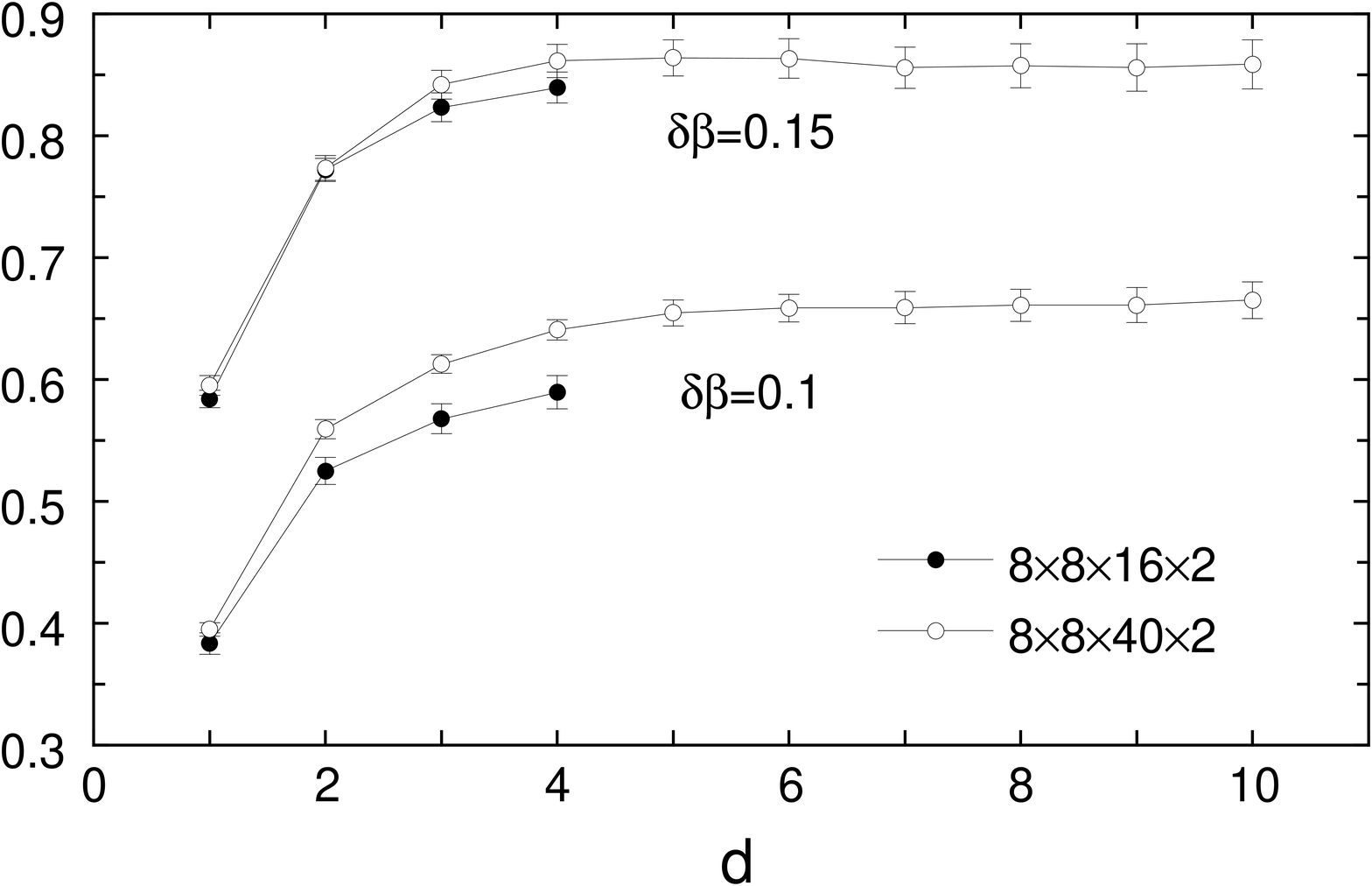}
\caption{{
$\tilde\sigma(d)$ defined as $\sigma/T_c^3$ with the operator method 
with restricting the summation in (\protect\ref{eq:operator}) 
to the regions within the distance $d$ from $z=0$ and $N_z/2$. 
($\tilde\sigma(N_z/4) \equiv \sigma/T_c^3$ given in 
Table~\protect\ref{tab:sigma-op}.)
%$\tilde\sigma$ is plotted for the $8^2 \times 16 \times 2$ and 
%$8^2 \times 40 \times 2$ lattices at $\delta\beta=0.1$ and 0.15.
}}
\protect\label{fig:sigma-op-width}
\end{figure}

%%%%%%%%%%%%%%%%%%%%%%%%%%%%%

\begin{figure}[p]
\epsfxsize=12cm \epsfbox{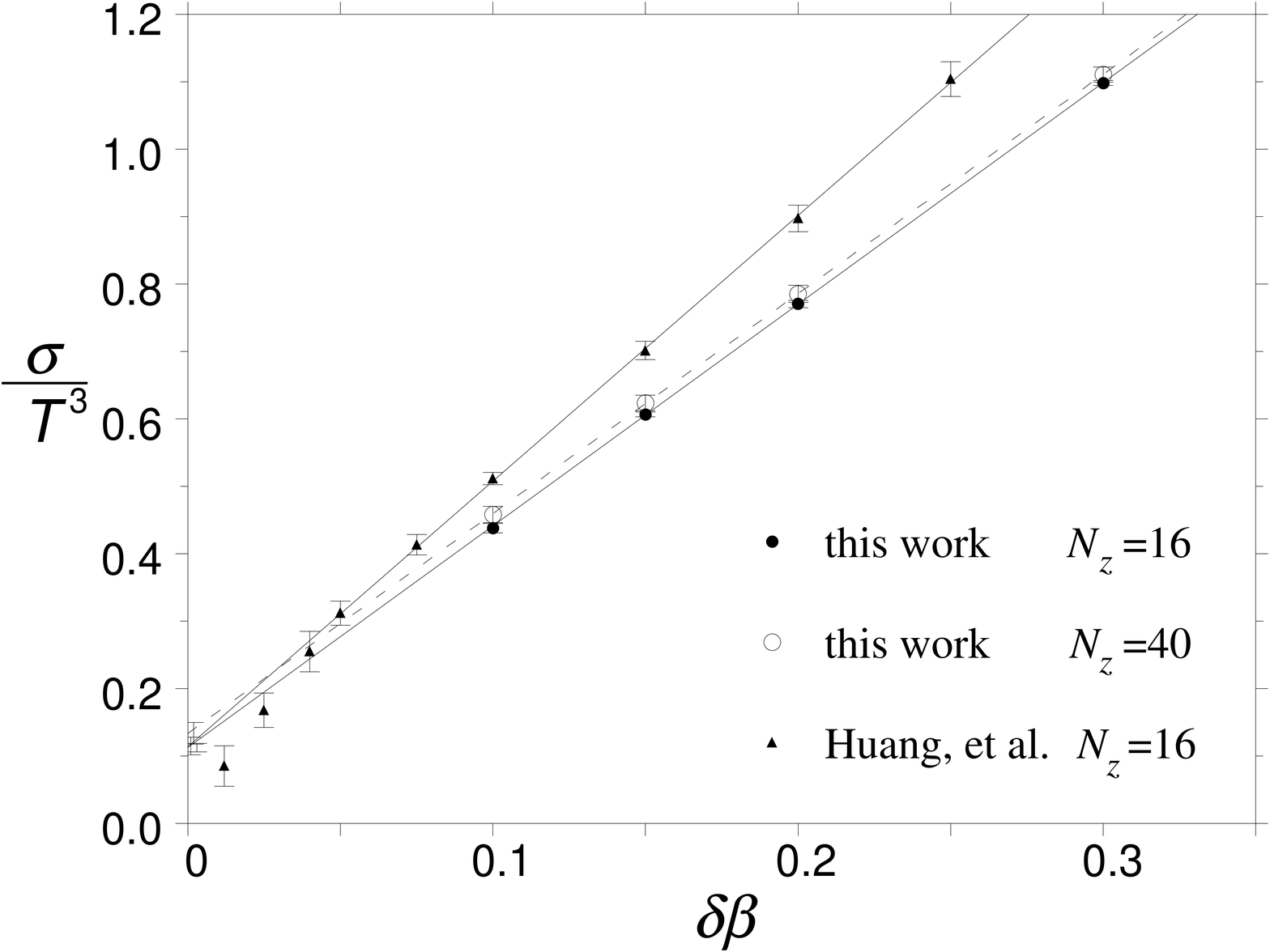}
\caption{{
$\sigma$ with the integral method as a function of $\delta\beta$. 
Filled circles are the results of 
the $8^2 \times 16 \times 2$ lattice, and open circles those of the
$8^2 \times 40 \times 2$ lattice. 
Triangles are the results of Huang, {\it et al.} \protect\cite{Boston90} %[5] 
for $N_z=16$.
}}
\protect\label{fig:sigma-int}
\end{figure}

%%%%%%%%%%%%%%%%%%%%%%%%%%%%%

\begin{figure}[p]
\epsfxsize=12cm \epsfbox{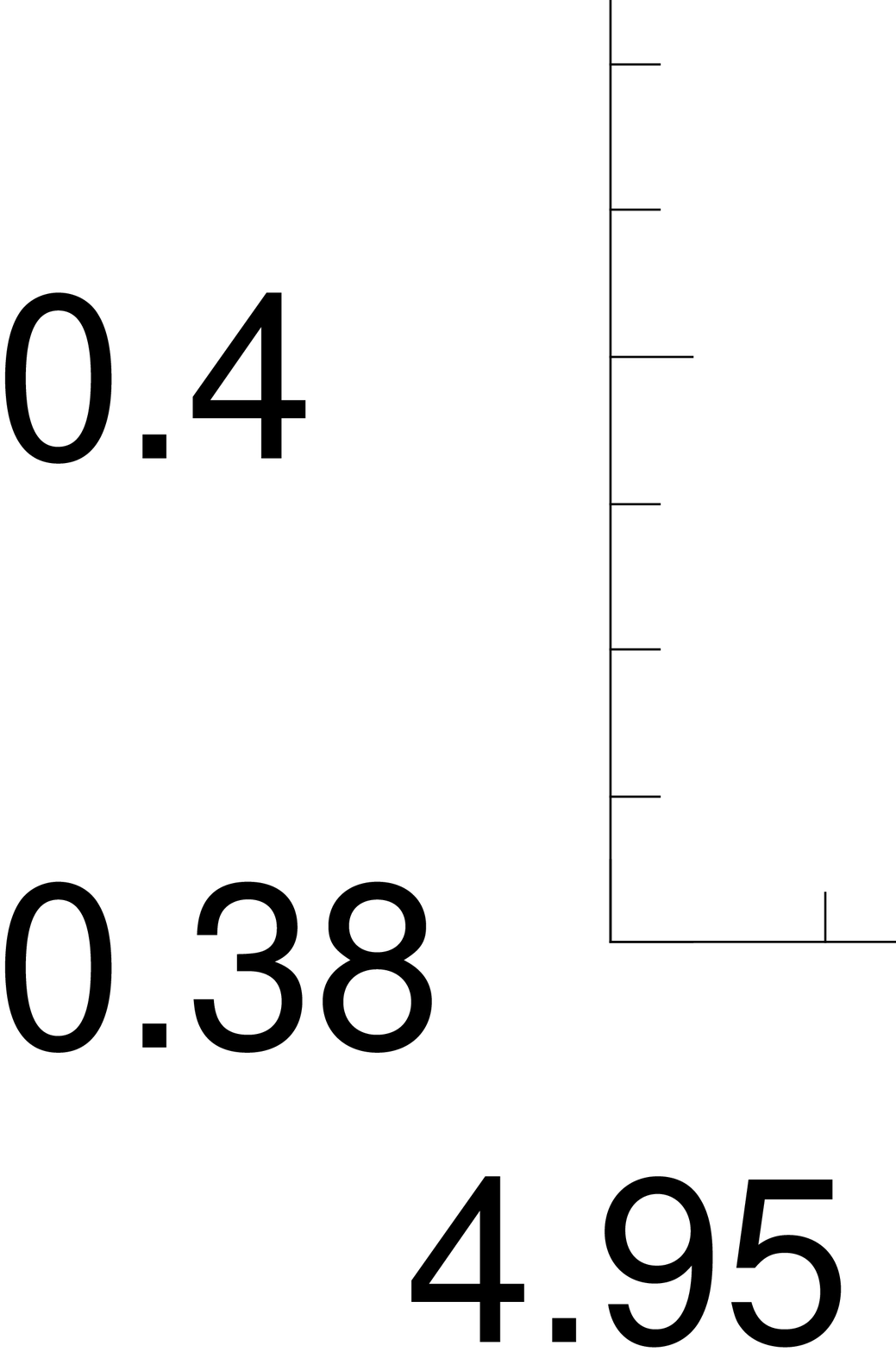}
\caption{{
Plaquette expectation value in domain 1 along the path CE of 
Fig.~\protect\ref{fig:intpath}
and that in domain 2 along BC on the $8^2 \times 16 \times 2$ lattice. 
Estimated errors are much smaller than the size of symbols. 
}}
\protect\label{fig:plaq-int16}
\end{figure}

%%%%%%%%%%%%%%%%%%%%%%%%%%%%%

\begin{figure}[p]
\epsfxsize=12cm \epsfbox{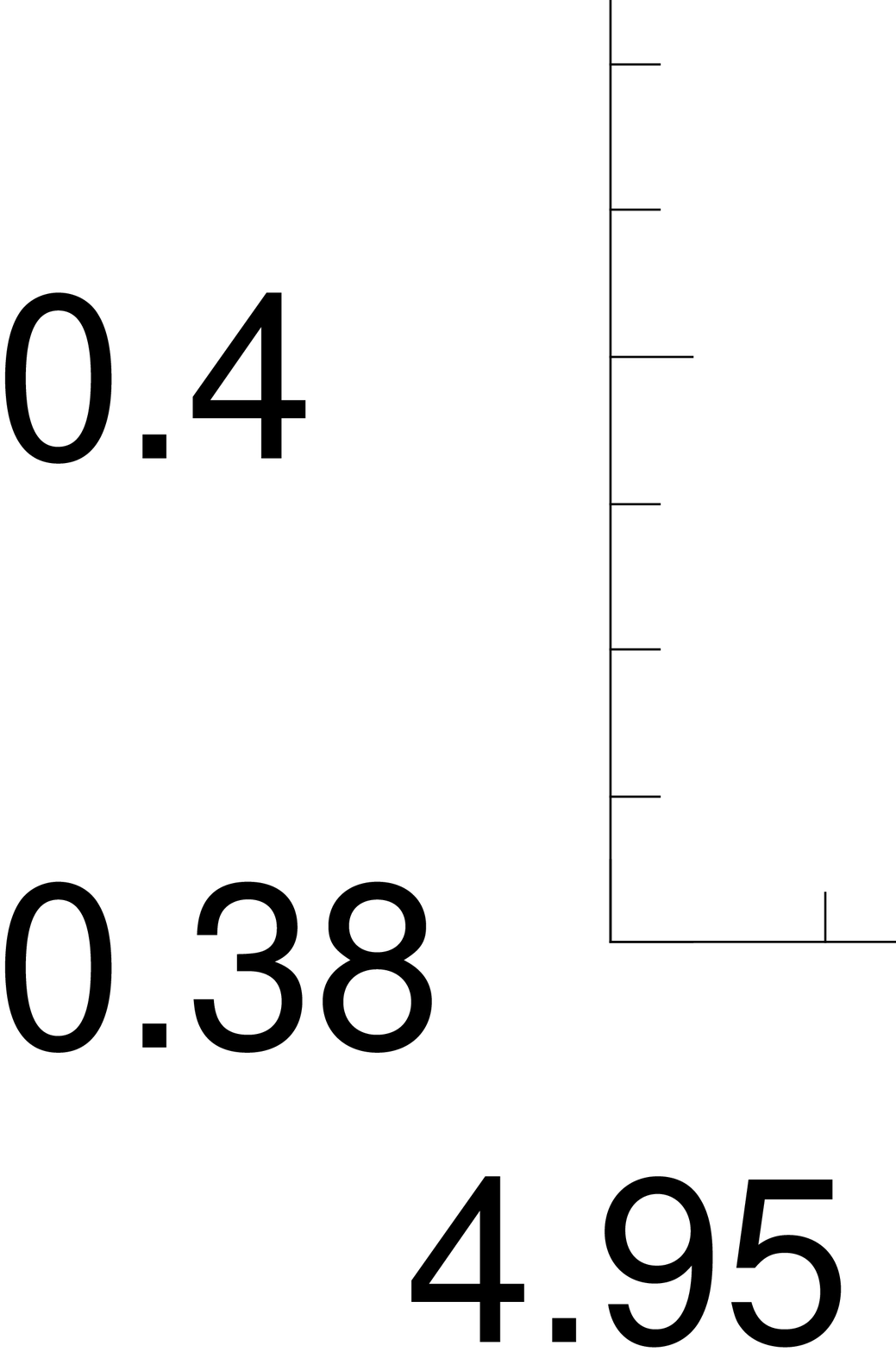}
\caption{{
The same as Fig.~\protect\ref{fig:plaq-int16} obtained 
on the $8^2 \times 40 \times 2$ lattice. 
}}
\protect\label{fig:plaq-int40}
\end{figure}

%%%%%%%%%%%%%%%%%%%%%%%%%%%%%

\begin{figure}[p]
\epsfxsize=12cm \epsfbox{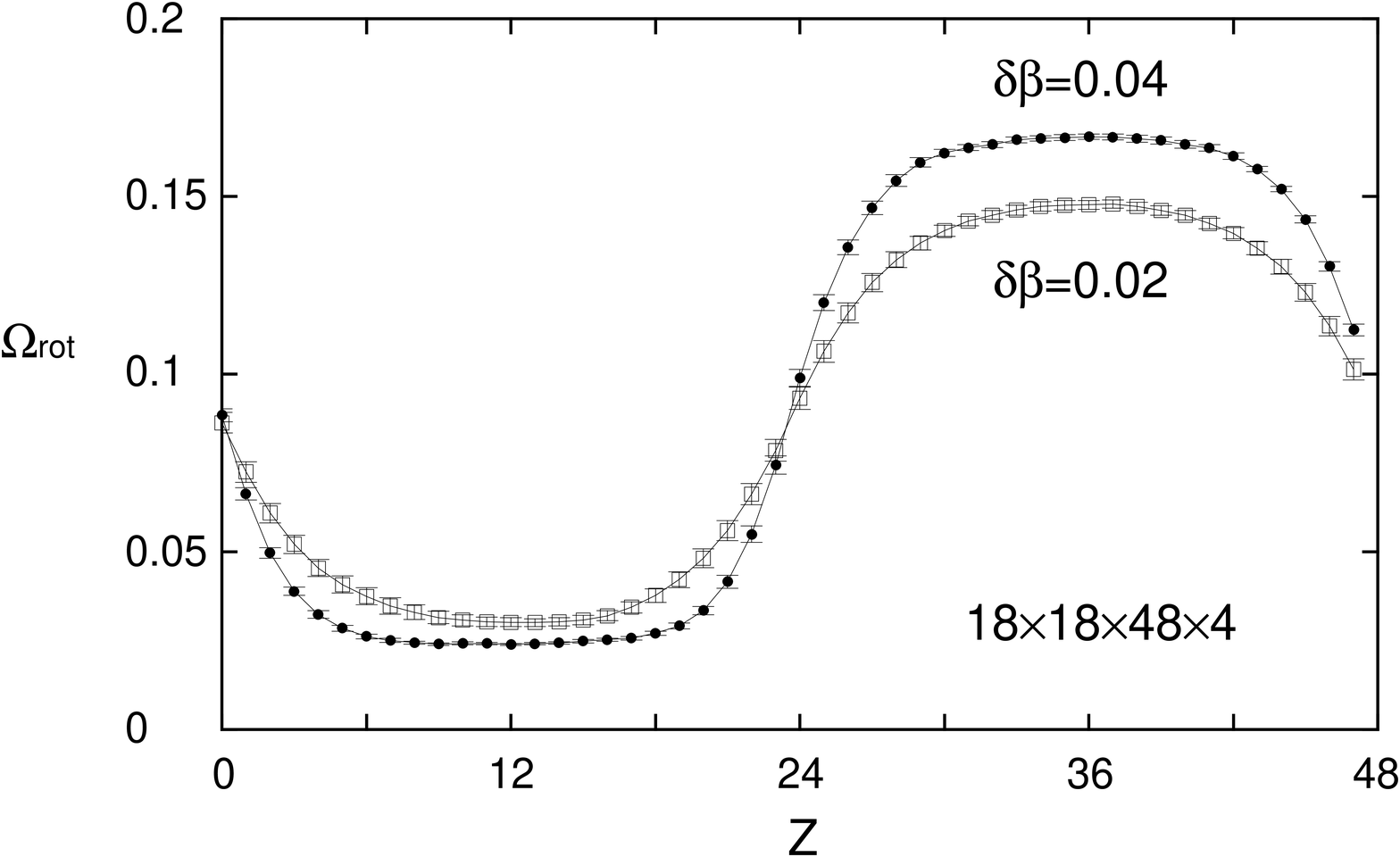}
\caption{{
Real part of the rotated Polyakov loop 
as a function of $z$ 
on the $18^2 \times 48 \times 4$ lattice for $\delta\beta=0.02$ 
and 0.04.
}}
\protect\label{fig:poly-T4}
\end{figure}

%%%%%%%%%%%%%%%%%%%%%%%%%%%%%

\begin{figure}[p]
\epsfxsize=12cm \epsfbox{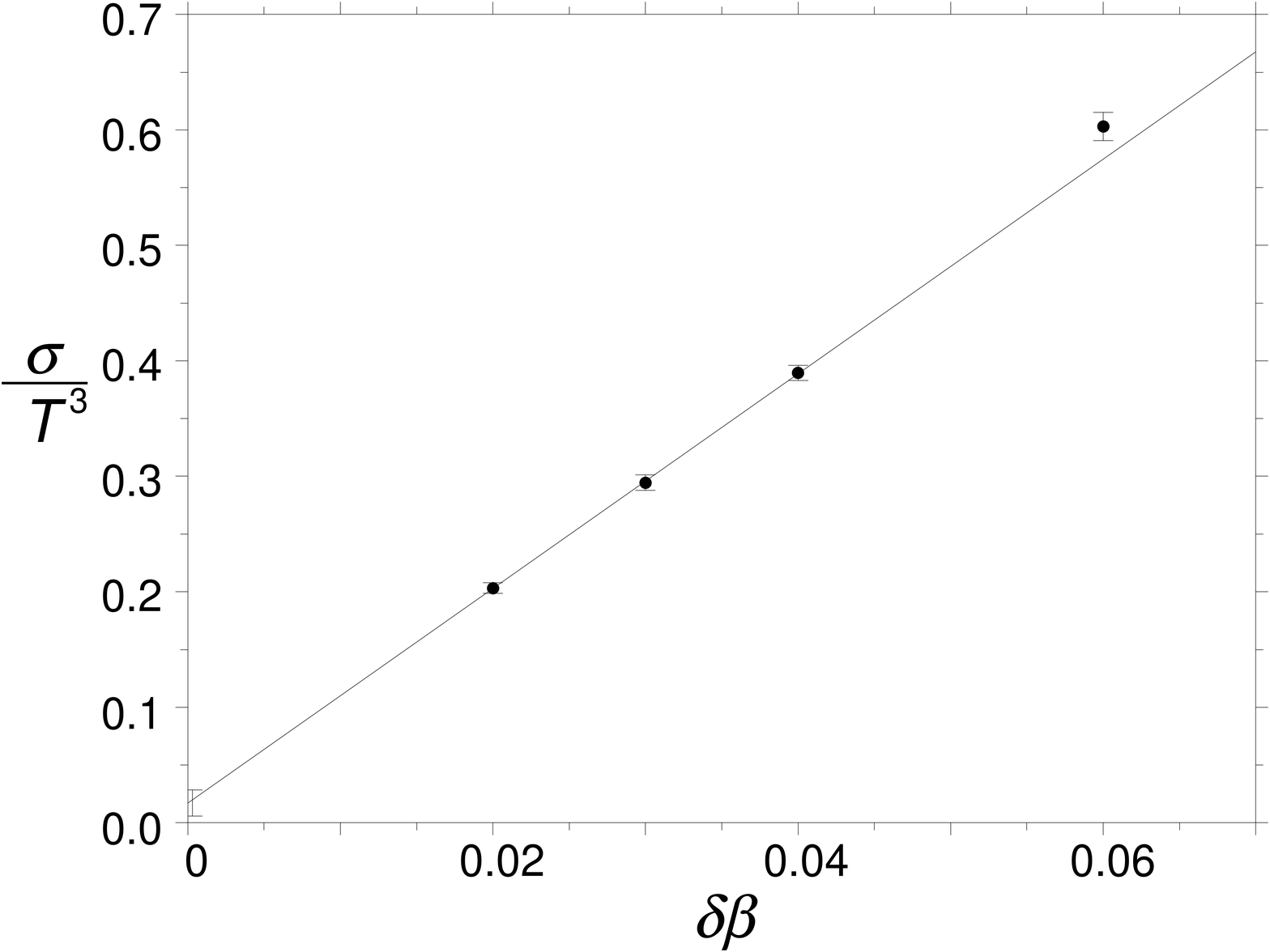}
\caption{{
$\sigma$ with the integral method as a function of $\delta\beta$ 
on the $18^2 \times 48 \times 4$ lattice. 
The line is the result of linear fit to the data at 
$\delta\beta=0.02$ -- 0.04.
}}
\protect\label{fig:sigma-intT4}
\end{figure}

\end{document}